\documentclass[12pt]{article}
\usepackage{epsfig}
\usepackage{comment}
\usepackage{latexsym}
\usepackage{color}
\newcommand{\mysquare}[0]{\raise-.2ex\hbox{{\Large$\Box$}}}
\def\lsim{\mathrel{\rlap {\raise.5ex\hbox{$ < $}}
{\lower.5ex\hbox{$\sim$}}}}
\def\gsim{\mathrel{\rlap {\raise.5ex\hbox{$ > $}}
{\lower.5ex\hbox{$\sim$}}}} \topmargin -1.5cm \textheight=22.5cm \textwidth=16.5cm
\catcode`\@=11
\newcount\hour
\newcount\minute
\newtoks\amorpm
\hour=\time\divide\hour by60 \minute=\time{\multiply\hour by60 \global\advance\minute by-\hour}
\edef\standardtime{{\ifnum\hour<12 \global\amorpm={am}%
        \else\global\amorpm={pm}\advance\hour by-12 \fi
        \ifnum\hour=0 \hour=12 \fi
        \number\hour:\ifnum\minute<10 0\fi\number\minute\the\amorpm}}
\edef\militarytime{\number\hour:\ifnum\minute<10 0\fi\number\minute}
\def\draftlabel#1{{\@bsphack\if@filesw {\let\thepage\relax
   \xdef\@gtempa{\write\@auxout{\string
      \newlabel{#1}{{\@currentlabel}{\thepage}}}}}\@gtempa
   \if@nobreak \ifvmode\nobreak\fi\fi\fi\@esphack}
        \gdef\@eqnlabel{#1}}
\def\@eqnlabel{}
\def\@vacuum{}
\def\draftmarginnote#1{\marginpar{\raggedright\scriptsize\tt#1}}
\def\draft{\oddsidemargin -.2truein
        \def\@oddfoot{\sl preliminary draft \hfil
        \rm\thepage\hfil\sl\today\quad\militarytime}
        \let\@evenfoot\@oddfoot \overfullrule 3pt
        \let\label=\draftlabel
        \let\marginnote=\draftmarginnote
   \def\@eqnnum{(\theequation)\rlap{\k

 ern\marginparsep\tt\@eqnlabel}%
\global\let\@eqnlabel\@vacuum}  }

\newcommand{\be}[0]{\begin{equation}}
\newcommand{\ee}[0]{\end{equation}}
\newcommand{\ba}[0]{\begin{eqnarray}}
\newcommand{\ea}[0]{\end{eqnarray}}

\def\bs{\begin{subequations}}
\def\es{\end{subequations}}

\def\thebibliography#1{%
\vskip 0.5cm \centerline{\bf \Large References}
\list{%
[\arabic{enumi}]}{\settowidth\labelwidth{[#1]} \leftmargin\labelwidth \advance\leftmargin\labelsep
\usecounter{enumi}}
\def\newblock{\hskip .11em plus .33em minus .07em}
\sloppy\clubpenalty4000\widowpenalty4000 \sfcode`\.=1000\relax}

\renewcommand{\theequation}{\arabic{section}.\arabic{equation}}

\renewcommand{\section}{\setcounter{equation}{0}\@startsection
{section}{1}{0mm}{-\baselineskip}{0.5\baselineskip} {\normalfont\Large\bfseries}}

\renewcommand{\subsection}{\@startsection
{subsection}{2}{0mm}{-\baselineskip}{0.5\baselineskip} {\normalfont\large\bfseries}}

\renewcommand{\subsubsection}{\@startsection
{subsubsection}{3}{0mm}{-\baselineskip}{0.5\baselineskip} {\normalfont\normalsize\slshape}}

\usepackage{amssymb,amsfonts}
\usepackage{graphicx}
\usepackage{cite}

\def\bc{\begin{center}}
\def\ec{\end{center}}
\def\bea{\begin{eqnarray}}
\def\eea{\end{eqnarray}}

\def\nn{\nonumber}

\def\k{\kappa}

\def\tt{\tilde t}
\def\hk{\hat k}

\def\and{\quad\mbox{and}\quad}

\newcommand{\Z}{\mathbb{Z}}

\begin{document}
\begin{titlepage}
\begin{flushright}
LPTENS--08/21, CPHT--RR008.0308, March 2008
\end{flushright}

\vspace{2mm}

\begin{centering}
{\bf\huge Thermal and Quantum}\\
\vspace{2mm}
{\bf\huge Superstring Cosmologies$^\ast$}\\

\vspace{6mm}
 {\large Tristan Catelin-Jullien$^{1}$, Costas~Kounnas$^{1}$\\
 Herv\'e~Partouche$^2$ and Nicolaos~Toumbas$^3$}

\vspace{4mm}

$^1$ Laboratoire de Physique Th\'eorique,
Ecole Normale Sup\'erieure,$^\dagger$ \\
24 rue Lhomond, F--75231 Paris Cedex 05, France\\
{\em  catelin@lpt.ens.fr, Costas.Kounnas@lpt.ens.fr}

\vskip .1cm

$^2$ Centre de Physique Th\'eorique, Ecole Polytechnique,$^\diamond$
\\
F--91128 Palaiseau, France\\
{\em Herve.Partouche@cpht.polytechnique.fr} \vskip .1cm

$^3$ Department of Physics, University of Cyprus,\\ Nicosia 1678, Cyprus \\ {\em nick@ucy.ac.cy}
 \vspace{6mm}

{\it Talk given by Costas Kounnas at the 3rd RTN Workshop Forces-Universe,\\
Valencia, Spain  October 1-5, 2007, \\
and by Herv\'e Partouche at the ``String Phenomenology and Dynamical Vacuum
Selection'' Workshop, Liverpool, UK, March 27-29, 2008.}
 \vspace{6mm}

{\bf\Large Abstract}

\end{centering}


\noindent We consider four dimensional heterotic string backgrounds for which supersymmetry is spontaneously broken via
the Scherk-Schwarz mechanism on an internal spatial cycle and by finite temperature effects.
We concentrate on initially flat backgrounds with
$N=4$ and $N=2$ amount of supersymmetry.   
Thermal and quantum corrections give rise to a non-trivial cosmological evolution. We show that these corrections are
under control and calculable due to the underlying no-scale structure of the effective supergravity theory. 
The effective 
Friedmann-Hubble equation involves 
a radiation term $\sim 1/a^4$ and a curvature term $\sim 1/a^2$, whose
coefficients are functions of ratio of the gravitino mass scale to the temperature.

\vspace{5pt} \vfill \hrule width 6.7cm \vskip.1mm{\small \small \small \noindent $^\ast$\ Research
partially supported by the EU (under the contracts MRTN-CT-2004-005104, MRTN-CT-2004-512194,
MRTN-CT-2004-503369, MEXT-CT-2003-509661), INTAS grant 03-51-6346,
 CNRS PICS 2530, 3059 and 3747,  and ANR (CNRS-USAR) contract
 05-BLAN-0079-01 (01/12/05).\\
$^\dagger$\ Unit{\'e} mixte  du CNRS et de l'Ecole Normale Sup{\'e}rieure associ\'ee \`a
l'universit\'e Pierre et Marie Curie (Paris
6), UMR 8549.\\
 $^\diamond$\ Unit{\'e} mixte du CNRS et de l'Ecole Polytechnique,
UMR 7644.}

\end{titlepage}
\newpage
\setcounter{footnote}{0}
\renewcommand{\thefootnote}{\arabic{footnote}}
 \setlength{\baselineskip}{.7cm} \setlength{\parskip}{.2cm}


\setcounter{section}{0}
\section{Introduction}

\noindent
A fundamental challenge for string theory is to explain the cosmology of our Universe. 
How can the theory describe, or even better predict, basic features of our Universe?
Despite considerable effort over the last few years (see \cite{ABS} -- \cite{Maldacena:2000mw} for
a partial list of 
references), 
still a concrete string theoretic framework for studying cosmology is lacking. The purpose of
this article is to report some progress toward this direction by exhibiting new, 
physically relevant cosmological solutions of superstring theory. These solutions  
were obtained and analyzed recently in \cite{ckpt}, after taking into account thermal and quantum
corrections in 
superstring models for which supersymmetry is spontaneously broken.

\noindent At the level of classical string compactifications (with or without fluxes), 
it seems difficult to obtain realistic, tractable cosmological solutions. 
In most cases, the classical ground
states correspond to static Anti-de Sitter or flat backgrounds and not to cosmological ones.
But this classical analysis neglects the thermal and quantum corrections, which inevitably must play
an
important role in any attempt to identify non-trivial cosmological states.
   
\noindent It is precisely this direction that we wish to explore in this article. It involves 
studying cosmologies that are generated dynamically at the quantum level of 
string theory \cite{KP,KPThermal,AK86,ckpt}.
For
certain cases the quantum and thermal corrections are under control
due to the very special structure of the underlying effective supergravity theory in its
spontaneously broken
supersymmetric phase. 

\noindent In order to see how cosmological solutions emerge naturally in this context, consider the
case of an initially
supersymmetric flat string background at finite temperature. The thermal fluctuations produce a 
calculable energy density whose 
back-reaction on the space-time metric and on certain moduli fields
gives rise to a cosmological evolution. For temperatures below the Hagedorn temperature, 
the evolution of the universe is known to be radiation dominated \cite{Matsuo:1986es,vafabrand}. 

\noindent More interesting cases are those where space-time supersymmetry is spontaneously broken at
the string
level via freely acting orbifolds \cite{Scherk:1978ta}--\cite{akd}. In these cases, 
the thermal and supersymmetry breaking couplings correspond to a
generalization of Scherk-Schwarz compactification in superstrings. 
are
The thermal corrections are implemented by introducing a coupling of the 
space-time fermion number
$Q_F$ to the string momentum and winding numbers associated to the Euclidean time cycle $S_T^1$.
The breaking of supersymmetry is generated by a similar coupling of an internal $R$-symmetry charge
$Q_R$ to the momentum and winding numbers associated to an internal spatial cycle
$S_M^1$, e.g. the $X_5$ coordinate cycle.

\noindent 
Two special mass scales appear both associated
with the breaking of supersymmetry: the temperature scale $T\sim 1/(2\pi R_0)$ and the
supersymmetry breaking scale $M\sim 1/(2\pi R_5)$, where $R_0$ and $R_5$ are the radii of the
Euclidean time
cycle, $S_T^1$, and of the internal spatial cycle, $S_M^1$, respectively. The initially degenerate
mass levels
of bosons and fermions split by an amount proportional to $T$ or $M$, according
to the charges $Q_F$ and $Q_R$. This mass splitting gives rise to a non-trivial free energy
density, 
which incorporates simultaneously the thermal
corrections and quantum corrections due to the supersymmetry breaking boundary conditions along the
spatial
cycle $S_M^1$.
The back-reaction on the initially flat
space-time metric results in deferent kinds of cosmological evolutions, depending on the
initial amount of supersymmetry $(N=4,N=2,N=1)$.

\noindent In \cite{ckpt} we concentrated on four dimensional heterotic models with initial $N=4$ and
$N=2$ amount of 
supersymmetry, leaving  
the phenomenologically more interesting $N=1$ cases for future work. Below we summarize some of our
main results. 

\section{Thermal and quantum corrections in heterotic backgrounds }
We study the class of four dimensional string backgrounds obtained by toroidal
compactification of the heterotic string on $T^6$ and $T^6/\Z_2$ orbifolds. The initial amount of
space-time supersymmetry is $N_4=4$ for the $T^6$ models and $N_4=2$
for the orbifold models. Space-time
supersymmetry is then spontaneously broken by introducing Scherk-Schwarz boundary conditions on an
internal spatial cycle and by thermal corrections. 

\noindent The four dimensional one-loop effective action in string frame is given by \be S=\int
d^4 x\sqrt{-\det g}\left( e^{-2\phi}({1 \over 2}R+2\partial_{\mu}\phi\partial^{\mu}\phi+\dots)-{\rm
\cal V}_{\rm String}\right),\ee where $\phi$ is the $4d$ dilaton field. The ellipses stand for
the kinetic terms of other moduli fields. At zero temperature, the
effective potential ${\rm \cal V}_{\rm String}$ is given in terms of the one-loop Euclidean string
partition function as follows: \be {Z \over V_4}= -{\rm \cal V}_{\rm String}\ee with $V_4$ the $4d$
Euclidean volume. 
At finite temperature, the one-loop Euclidean partition function determines the free energy density
and pressure: \be {Z \over V_4}= -{ \cal F}_{\rm String}=P_{\rm String}.\ee 

\noindent In order to determine the back-reaction on the metric and on certain moduli fields, it is
convenient to work in the Einstein frame. For this purpose, we define the complex field $S$, 
$S=e^{-2\phi}+i\chi$, where $\chi$
is the axion field. Then after the Einstein rescaling of the metric, the one loop effective action
becomes: \be S=\int d^4x \sqrt{-\det g}\left[{1\over 2}R -g^{\mu\nu}~K_{I \bar J}
 ~\partial_{\mu}\Phi_I \partial_{\nu} \bar\Phi_{\bar J}~
-{1\over s^2}~{\rm \cal V}_{\rm String} (\Phi_I, \bar\Phi_{\bar I})\right]. \ee Here  $K_{i \bar
\jmath}$ is the metric on the scalar field manifold $\{\Phi_I\}$, parameterized by various
compactification moduli and the field $S$. 
This manifold includes also the main moduli fields
$T_I,\, U_I,~I=1,2,3$, which are the volume and complex structure moduli of the three internal
$2$-cycles respectively. 

\noindent In the Einstein frame the effective potential is rescaled by a factor $1/ s^2$, where
$s={\rm Re} S=e^{-2\phi}$. 
We have ${\rm \cal V}_{\rm Ein}={\rm \cal V}_{\rm String}/s^2$.
We always work in
gravitational mass units, with $ M_G ={1\over \sqrt{ 8 \pi G_N}}=2.4\times10^{18}$ GeV.

\noindent What will be crucial in our analysis are some fundamental scaling properties of ${\rm \cal
V}_{\rm Ein}$ in the limit of
large $R_0,R_5\gg 1$. In this limit, only the temperature scale $T\sim 1/\sqrt{s}R_0$ and three of
the main moduli fields,
$\{S,T_1,U_1\}$ appear in ${\rm \cal V}_{\rm Ein}$. All other moduli appear in exponentially
suppressed contributions:
\be{\rm
\cal V}_{\rm Ein}~\simeq ~{F\left(~{sR_0^2 \over sR_5^2},\dots\right)\over (st_1u_1)^2}~+~ {\rm \cal
O}(e^{-c_0R_0-c_5R_5}),\ee
where the function $F$ will be determined later on.
Freezing all other moduli, the classical K\"alher potential takes
a no-scale structure \cite{noScale}, as was expected from the effective field
theory approach: 
\be K=-\log ~(S+\bar S) -\log ~(T_1+\bar
T_1)- \log ~(U_1+\bar U_1)\equiv -3\log ~(Z+\bar
Z),
\ee
with $z={\rm Re}\, Z$ and $z^3=st_1u_1$.

\noindent The classical superpotential is constant, and so the
gravitino mass scale is given by 
\be
M^2=8e^K= {1\over st_1u_1}={1\over
z^3}.\ee
Freezing further ${\rm Im} Z$ and defining the
field $\Phi$ by 
\be e^{2\alpha \Phi}=
M^2={8\over (Z+\bar Z)^3},\ee we obtain the following kinetic term 
\be - g_{\mu\nu}~3{\partial_{\mu} Z\partial_{\nu} \bar Z \over (Z+\bar
Z)^2}= -g_{\mu\nu}~{\alpha^2 \over 3}~\partial_{\mu}
\Phi\partial_{\nu} \Phi~ \ee from the K\"ahler potential. The choice $\alpha^2=3/2$
 normalizes canonically the kinetic term of the
no-scale  modulus $\Phi$.

\noindent
In all, the effective potential in Einstein frame acquires the following structure:
\be 
{\rm \cal V}_{\rm Ein}\simeq M^4~
F\left(~{sR^2_0\over sR^2_5}, \dots\right)\simeq
M^4~F\left({M^2\over T^2},
{m^2_Y\over T^2}\right).\ee
 The possible dependence on
other Susy mass scales  $M^2_Y$ will become
clear latter on, when we consider explicit examples.

\section{Thermal and spontaneous breaking of supersymmetry}

\noindent We first consider the case of a heterotic string
background with maximal space-time supersymmetry ($N_4=4$).
All nine spatial directions as well as the
Euclidean time are compactified on a ten dimensional torus.
At zero temperature and in the absence of Susy breaking couplings, the Euclidean
string partition function is zero due to space-time supersymmetry. 

\noindent At finite temperature and in the presence of a Scherk-Schwarz Susy breaking coupling,
the result is a well defined finite quantity \cite{Atick:1988si}-\cite{akd}. At genus one the string
partition function is
given by: 
$$
Z=\int_F ~{d\tau d\bar{\tau}\over 4{\rm
Im}\tau}~{1\over 2}\sum_{a,b}~~
{(-)^{a+b+ab}~\theta\left[^a_b\right]^4\over\eta(\tau)^{12}~
{\bar\eta}(\bar\tau)^{24}}~~\Gamma_{(5,21)}(R_I)
~\Gamma_{(3,3)}(R)
$$
\be
\times~\sum_{h_0,g_0}~\Gamma\left[^{h_0}_{g_0}\right](R_0)~
(-)^{g_0 a+h_0b +g_0h_0}
~\sum_{h_5,g_5}~~\Gamma\left[^{h_5}_{g_5}\right](R_5)~
(-)^{g_5 a+h_5b +g_5h_5}~.
\ee
The non-vanishing of the partition function is due
to the non-trivial coupling of the $\Gamma(R_0)$ and the $\Gamma(R_5)$ shifted lattices to the spin
structures $(a,b)$. 
Here, the argument 
$a$
is zero for space-time bosons and one for space-time fermions.
The shifted lattices are given by
$
\Gamma_{(1,1)}\left[^h_g\right](R)= \sum_{m,n}R({\rm Im}\tau)^{-{1\over2}}~e^{-\pi
R^2{|2m+g+(2n+h)\tau|^2\over {\rm Im} \tau}}~$.
We are interested in the
case for which the radii of three spatial directions are very large, $R_x=R_y=R_z\equiv R\gg 1$, so
that 
the three dimensional spatial volume factorizes
$\Gamma_{(3,3)}\cong R^{3}~{{\rm
Im}\tau}^{-{3\over 2}}= {(V_3/ (2\pi)^3)}~{{\rm Im}\tau}^{-{3\over 2}}$. 

\noindent Before we proceed, the following comments are in order:
\begin{itemize}
\item 
The sector $(h_0,g_0)=(h_5,g_5)=(0,0)$ gives zero contribution due to the fact that we started
with a supersymmetric background.
 \item 
 In the odd winding sectors, $h_0=1$ and/or $h_5=1$, the partition function
diverges when $R_0$ and/or $R_5$ are between the Hagedorn radius $R_H=(\sqrt{2}+1)/2$ and its dual
$1/R_H$:
${1\over R_H} < R_{0,5} < R_H $. 
The divergence is due to winding states that become tachyonic. Their condensation drives the system
towards a 
phase transition \cite{Atick:1988si}-\cite{akd}.
\item
In the regime $R_0,\, R_5\gg 1$, there are
no tachyons. As we will see, the odd winding sectors as well as the
string oscillator states give exponentially
suppressed contributions to the partition function.
The contributions of the internal
$\Gamma_{(5,21)}(R_I)$ lattice states are also
exponentially suppressed, provided that
the moduli $R_I$ are of order unity.
\end{itemize}


\noindent Thus for large $R_0, \, R_5$, only sectors for which $h_0=h_5=0$ 
contribute significantly. By utilizing Jacobi identities
involving the theta functions, we can see that when $h_0=h_5=0$, we get a non-zero contribution only
if
$g_0+g_5=1$. 

\noindent Next, using
the relation $ \Gamma(R)=\Gamma\left[^0_0\right]+\Gamma\left[^0_1\right]
+\Gamma\left[^1_0\right]+\Gamma\left[^1_1\right]$ and neglecting the odd winding 
sectors, we may replace \be \Gamma\left[^0_1\right]\rightarrow
\Gamma(R)-\Gamma\left[^0_0\right]=\Gamma(R)-{1\over 2}\Gamma(2R)\ee
in the integral expression for $Z$. For each lattice term we decompose the contribution in modular
orbits: $(m_i,n_i)=(0,0)$ and $(m_i,n_i)\neq(0,0)$. For $(m_i,n_i)\neq(0,0)$, the integration over
the
fundamental domain is equivalent with the integration over the whole strip but with $n_i=0$. Notice
also that the $(0,0)$ contribution of
$\Gamma(R)$ cancels the one of ${1\over 2}\Gamma(2R)$. We are left with the following
integration over the whole strip: 
\be Z = ~{V_5 \over (2\pi)^5}\sum_{g_0+g_5=1} \int_{||} {d\tau d\bar{\tau}\over
~{4\rm Im}\tau^{7\over 2}~}~{\theta\left[^{1}_{0}
\right]^4 \Gamma_{(5,21)}(R_I)\over \eta(\tau)^{12}~ {\bar\eta}(\bar\tau)^{24}}
\sum_{m_0,m_5}~e^{-\pi R_0^2{(2m_0+g_0)^2\over {\rm Im} \tau}}~e^{-\pi R_5^2{(2m_5+g_5)^2\over {\rm
Im} \tau}}.\ee

\noindent The integral over $\tau_1$ imposes the left-right level matching condition. The
left-moving part
contains the ratio $ {\theta\left[^{1}_{0} \right]^4/ \eta^{12}} =2^4+{\rm \cal
O}(e^{-\pi\tau_2})$, which implies that the lowest contribution is at the massless level. Thus
after the integration over $\tau_1$ $(\tau_2\equiv t)$, the partition function takes the form 
\be Z
=2^4D_0~{V_5 \over (2\pi)^5}~\sum_{g_1,g_2}{\left(1-(-)^{g_1+g_2}\right)\over 2}~ \int_0^{\infty
}{dt \over 2t^{7 \over 2}~}\sum_{m_1,m_2}~e^{-\pi R_0^2{(2m_1+g_1)^2\over t}-\pi
R_5^2{(2m_2+g_2)^2\over t}},\label{partitionmixed}  \ee 
up to exponentially suppressed contributions that we drop.
The factor $2^4~ D_0$ is the multiplicity of the massless level. 

\noindent Changing the integration variable by setting $t=\pi
(R_0^2 (2m_1 +g_1)^2~+~R_5^2 (2m_2 +g_2)^2)~x$, the integral over $x$ can be expressed neatly in
terms of 
Eisenstein functions of order $k=5/2$:
\be E_k(U) =
\sum_{(m,n)\neq(0,0)}\left(\frac {{\rm Im}\ U}{|m+nU|^2}\right)^k.\ee
The pressure in the
Einstein frame can be written as \be P={Z\over V_4} = ~C_T~{T^4}~ f_{5/2}(u)+C_V~
M^4~ { f_{5/2}(1/u) \over u} \label{pressuremixed}, \ee
where $u=R_0/R_5=M/T$, and 
2
\be f_k(u) = u^{k-1}\left(\frac 1{2^{k}}E_{k}\left({iu\over 2}\right) - \frac
1{2^{2k}}E_{k}(iu)\right).\ee


Here $C_T=C_V \sim n^*$, where $n^*=8D_0$ is the number of massless fermion/boson pairs.
In this particular
model the coefficients $C_T$ and $C_V$ are equal due to the underlying
gravitino mass/temperature duality. For fixed $u$ the first term stands for the thermal
contribution to the pressure while the second term stands for minus the effective potential.

\noindent We conclude this section with some further comments.
\begin{itemize}
\item
The coefficient $C_T$ is fixed and positive as
it is determined by the number of all massless
boson/fermion pairs in the initially supersymmetric theory.
\item
The coefficient $C_V^{\rm }$ will depend on the way
the Susy-breaking operator $Q_R$ couples to the left
and right movers.
In general, $Q_F\neq Q_R$ and the temperature /
gravitino mass duality will be broken. Then $C_V$ can be either
positive or negative.
\item
For the $T_6/\Z_2$ orbifold models with $N=2$ initial supersymmetry, and with $Q_R$
acting only on the left-movers such that $Q_R\neq
Q_F$, the net contribution of the twisted
sectors to $C_V$ is negative \cite{ckpt}.
The change of sign indicates that in the
twisted sectors, the states that become massive are the
bosons rather than the fermions.
\end{itemize}

\subsection{Small mass scales from Wilson line deformations}
A generic supersymmetric heterotic background may contain in its spectrum massive super-multiplets
whose mass is obtained by switching on non-trivial continuous Wilson-lines \cite{Narain:1985jj}. 
This is a stringy
realization of the Higgs mechanism, breaking spontaneously the initial gauge group to smaller
subgroups. 

\noindent We restrict to arbitrary and small Wilson line deformations starting from a given
supersymmetric background where $R_I,\, I=6, 7, \dots, 10$ are of the order the string scale. 
 In the zero
winding sector, a Wilson line just modifies the Kaluza-Klein momenta, and the corresponding
Kaluza-Klein mass becomes 
\be {m_I^2\over R_I^2}~~\longrightarrow~~  {(m_I+~y_I^a~Q_a)^2\over
R_I^2},
\ee 
where $Q_a$ is the charge operator associated to the Wilson-line $y_I^a$. We can distinguish
two different cases:  $I=5$ where $R_5$ is large, and
$I=6,\dots,10$ where the $R_I$ are of order the string scale.

\noindent Here we shall consider the second case $I=6, 7, \dots, 10$. In this case, we can set the
momentum and 
winding numbers to zero,
$m_I=n_I=0$, so that the relevant modification in the partition function is the insertion of the
term:
\be e^{-\pi t \left({y_I^a Q_a\over R_I}\right)^2}~\simeq ~1~{-\pi t
\left({y_I^a Q_a \over R_I}\right)^2}.\ee
Then incorporating the effects of the Wilson lines up to quadratic order, we get for the overall
pressure:
\be
P =C_T T^4~f_{5\over 2}(u)~-~D_T
~T^2~M_Y^2~f_{3\over 2}(u)~
+~C_V ~M^4~{f_{5\over 2}(1/u)\over
u}~-~D_V~M^2~M_Y^2~{f_{3\over 2}(1/u)\over u}.
\ee
Here, $M_Y\sim
y_I^aQ_a/R_I$ introduces a new mass scale in the theory, which is qualitatively different than $T$
and $M$.
$M_Y$ is a supersymmetric mass scale rather than a Susy-breaking scale like $T$ and
$M$.

\subsection{Scaling properties of the thermal effective potential}
The final expression for $P$ contains three mass
scales: $M$, $T$ and $M_Y$. The first identity it satisfies follows from its
definition: 
\be \left(T{\partial \over
\partial T} +M{\partial \over
\partial M}+M_Y{\partial \over \partial M_Y}\right)~P=4P,
\ee
which can be best seen by
writing $P$ as 
\be P\equiv T^4~p_4(u)~
+~T^2~M_Y^2~p_2(u)=P_4+ P_2,~~u={M\over
T}.\ee
Using standard thermodynamic identities, we
can obtain the energy density $\rho=\rho_4+\rho_2$:
\be\rho \equiv T{\partial \over
\partial T}P-P = \rho_4+\rho_2\ee
with
 \be \rho_4=\left(3P_4-u{\partial \over
\partial u}P_4\right)~~~~~~
\rho_2 =\left(P_2-u{\partial \over
\partial u}P_2\right).\ee

\noindent In the sequel,
we allow the Susy-breaking scales $T$ and $M$ to vary with time while fixing the supersymmetric
mass scale $M_Y$ and also $u$, and investigate the back-reaction to the initially flat metric and
moduli fields.


\section{Gravitational equations and critical solution}
\noindent We assume that the back-reacted space-time metric is
homogeneous and isotropic 
\be
ds^2=-dt^2+a(t)^2~d\Omega_k^2,~~~~H=\left({\dot{a}\over a}\right),
\ee where $\Omega_k$ denotes the three dimensional space with
constant curvature $k$ and $H$ is the Hubble parameter.

\noindent From the fact that $-P$ plays the role of the effective
potential
and the relation between the gravitino mass scale $M$ and the no
scale modulus $\Phi$,
$M=e^{\alpha\Phi}$,
we obtain the field equation: \be
\ddot{\Phi}+3H\dot{\Phi}={\partial P\over \partial \Phi}=\alpha
u\left({\partial P \over \partial u}\right)_T =-\alpha\left(
\rho_4-3P_4 +\rho_2-P_2\right). 
\ee 

\noindent 
The remaining equations are the gravitational field equations. These are the
Friedmann-Hubble equation, 
\be 
\label{Hubble} 3H^2={1\over
2}\dot\Phi^2+\rho-{ 3 k\over a^2}, \ee
and the equation that follows from varying with respect to the
spatial components of the metric: \be 2\dot H +3H^2 =-{k\over
a^2}-P-{1\over2} \dot\Phi^2\, .
\label{vara} \ee This last equation can be replaced by the linear sum of the two gravitational field
equations, 
so that the kinetic term of $\Phi$ drops
out: \be \label{gravityeq} \dot H +3H^2 =-{2k\over a^2}+{1\over
2}(\rho-P)\, . \ee

\subsection{Critical solution}

The scaling
properties of the thermal effective potential suggest to search for a solution where all
varying mass scales of the system, $M(\Phi)$, $T$ and $1/ a$,
remain proportional during time evolution: \be \label{anzats}
e^{\alpha\Phi} \equiv M(\Phi)={1\over {\gamma}a}~~\Longrightarrow
~~H=-{\alpha} \dot{\Phi},~~~~ M(\Phi)\,= u \,T \, \ee with $\gamma$
and $u$ fixed in time. Our aim is thus to determine the constants
$\gamma$ and $u$.

\noindent Along the critical trajectory, the compatibility of the $\Phi$-equation of motion
with the gravitational field equations requires that

\be 
\label{equ} r_4= {6\alpha^2-1\over
2\alpha^2-1}p_4,~~~\left(r_4=~4p_4 ~~~~{\rm
for}~~\alpha^2={3\over2}\right) , \ee \be \label{eqk}
-2k\gamma^{2}~=~{2\alpha^2-1 \over 2}~{r_2-p_2\over
u^2}~M_Y^2,~~~\left( -2k\gamma^{2}={(r_2-p_2)\over u^2}~M_Y^2~~~~{\rm
for}~~\alpha^2={3\over 2}\right), 
\ee 
where $r_4=\rho_4/T^4$ and $r_2=\rho_2/(T^2\, M_Y^2)$.
The first equation is an algebraic equation for the complex structure-like ratio $u$. The second
equation determines the spatial curvature of the solution.

\noindent Having solved for the compatibility equations, the dynamics for the scale factor $a$ is
governed
by an effective Friedmann-Hubble equation as follows:
 \be
\label{cosmoeff} 3H^2= -{ 3\hat k\over a^2} +{c_r \over
a^4}\, , 
\ee 
where 
\be \nn  3\hat k=-{1\over
\gamma^2} ~{6\alpha^2\over 6\alpha^2-1}~{1\over
u^2}\left(~{3(2\alpha^2-1)\over 4}(r_2-p_2)+ r_2~\right)\,M_Y^2\, ,\ee 
\be
\label{CReff} c_r={1\over \gamma^{4}}~{6\alpha^2\over
6\alpha^2-1}~{r_4\over u^4}\,= {1\over \gamma^{4}}~{6\alpha^2\over
2\alpha^2-1}~{p_4\over u^4}\, . \ee 
The following comments are in order:
\begin{itemize}
\item
Clearly, a necessary condition for
the curvature $\hk$ not to vanish is to have non trivial Wilson
lines in any of the directions  6, 7, 8, 9, 10. Models with both positive and negative $\hat k$ can
be constructed \cite{ckpt}.
\item
The value of the ratio $u=M/T$
was obtained by solving the compatibility equations numerically.
It can be large or small depending on the model. In other words there are models with a 
hierarchy for the Susy-breaking scales $M$ and $T$.
In all models considered in \cite{ckpt}, the value for the effective coefficient $c_r$ was positive.
\end{itemize}

\section{Concluding remark}
The purpose of this talk is to emphasize the plausible existence of cosmological superstring
solutions, inflationary or not,
which are generated dynamically at the quantum sting level. Such cosmologies arise naturally from an
initially flat 
spacetime, once supersymmetry is spontaneously broken by thermal and quantum effects. 
They are examples of no-scale, radiatively induced cosmologies.
We believe that this new set-up will result in a coherent and fruitful framework in order to
understand superstring cosmology.

\section*{Acknowledgements}

 The work of C.K. and H.P. is partially supported by the EU contract MRTN-CT-2004-005104
and the ANR (CNRS-USAR) contract  05-BLAN-0079-01 (01/12/05). N.T. and C.K. are supported by the EU
contract MRTN-CT-2004-512194. H.P. is also supported by the EU contracts MRTN-CT-2004-503369 and
MEXT-CT-2003-509661, INTAS grant 03-51-6346, and CNRS PICS 2530, 3059 and 3747, while N.T. is also
supported by an INTERREG IIIA Crete/Cyprus program.

\end{document}